# Superconducting $Bi_2Te$: pressure-induced universality in the $(Bi_2)_m(Bi_2Te_3)_n$ series

Ryan L. Stillwell,[1] Zsolt Jenei,[2] Samuel T. Weir,[2] Yogesh K. Vohra,[3] Jason R. Jeffries,[1]

[1]Materials Science Division, Lawrence Livermore National Laboratory, Livermore, California 94550, USA

[2]Physics Division, Lawrence Livermore National Laboratory, Livermore, California 94550, USA

[3]Department of Physics, University of Alabama at Birmingham, Birmingham, Alabama 35294, USA

Using high-pressure magnetotransport techniques we have discovered superconductivity in $Bi_2Te$, a member of the infinitely adaptive $(Bi_2)_m(Bi_2Te_3)_n$ series, whose end members, Bi and $Bi_2Te_3$, can be tuned to display topological surface states or superconductivity. $Bi_2Te$ has a maximum $T_c$= 8.6 K at P= 14.5 GPa and goes through multiple high pressure phase transitions, ultimately collapsing into a bcc structure that suggests a universal behavior across the series. High-pressure magnetoresistance and Hall measurements suggest a semi-metal to metal transition near 5.4 GPa, which accompanies the hexagonal to intermediate phase transition seen via x-ray diffraction measurements. In addition, the linearity of $H_{c2}(T)$ exceeds the Werthamer-Helfand-Hohenberg limit, even in the extreme spin-orbit scattering limit, yet is consistent with other strong spin-orbit materials. Considering these results in combination with similar reports on strong spin-orbit scattering materials seen in the literature, we suggest the need for a new theory that can address the unconventional nature of their superconducting states.

## Introduction

The last several years have seen a large volume of research dedicated to materials whose properties are driven by the topology of their quantum states[1]. These states realize long theorized exotic quantum mechanical effects, such as Weyl fermions, predicted nearly a century ago, [2, 3] and others, like topological insulators (TI) and topological superconductors, only a few years ago[4, 5]. One of the more exciting prospects of TIs is that with proximity induced superconductivity[6] they provide the $Z_2$ non-trivial state allowing for promising applications in spintronics and quantum computation,[7] as well as the experimental realization of Majorana fermions[8].

One of the key properties necessary for making a TI is strong spin-orbit coupling, first discovered in HgTe quantum wells, and later found in several bismuth-based chalcogenides[9-12]. A particularly interesting family of these materials is the infinitely adaptive series $(A_2)_m(A_2Q_3)_n$, with A=(Bi, Sb) and Q=(Se, Te). Though members of this series are well known as some of the best thermoelectric materials, they were recently found to be topological insulators as well[13-16]. In particular, the end members of the infinitely adaptive $(Bi_2)_m(Bi_2Te_3)_n$ series, Bi and $Bi_2Te_3$, can be experimentally tuned to display topological surface states or superconductivity under appropriate conditions[11, 17-19].

In our search for functional materials that possess these exotic properties we synthesized and studied $Bi_2Te$, a member of the $(Bi_2)_m(Bi_2Te_3)_n$ series, with $m=2$ and $n=1$. Containing a Bi fraction of 0.67, $Bi_2Te$ is on the Bi rich side of this series, which spans the spectrum from pure Bi ($m=3$:$n=0$) to $Bi_2Te_3$ ($m=0$:$n=3$). Similar to other members of this series[20-22], $Bi_2Te$ transforms from an ambient pressure semimetal to a high pressure superconductor, progressing through multiple structural and electronic phases, including a high pressure body centered cubic structure with a maximum $T_c$ = 8.6 K at 14.5 GPa. This suggests a universal behavior seen across the $(Bi_2)_m(Bi_2Te_3)_n$ homologous series, as well as in elemental Te[23], [24], Se [25], and $Bi_2Se_3$ [26, 27].

## Experiment

Single crystals of $Bi_2Te$ were grown using the self-flux method. Stoichiometric amounts of elemental Bi and Te (99.999%, ESPI Metals) were combined in an alumina crucible and sealed under 500 mbar of high purity argon gas in a quartz tube and melted at 420 °C for 2 days. The solution was then cooled to 318 °C over a period 4 days, at which point the quartz tube was removed from the oven and centrifuged to remove excess flux. Large platelets were removed and ground into a coarse powder using a mortar and pestle. The coarse powder was then placed on a glass slide and chopped with a razor blade into a fine powder for powder x-ray diffraction characterization, which confirmed that they were single phase $Bi_2Te$.

For structural studies under pressure we used a conventional, membrane-driven diamond anvil cell (DAC) with 300μm diamond culets. The powdered $Bi_2Te$, 3-5μm particle size, was loaded into a 130-μm diameter sample chamber that was drilled out of a rhenium gasket preindented to 25 GPa, corresponding to an initial thickness of 23 μm. The chamber was also loaded with, fine copper powder (3-6$\mu$m, Alfa Aesar) as the pressure calibrant and neon precompressed to 30,000 psi as the pressure-transmitting medium. Room-temperature, angle-dispersive x-ray diffraction experiments were performed at HPCAT (beamline 16 BM-D) at the Advanced Photon Source at Argonne National Laboratory. The sample was illuminated with a 29.2 keV monochromatic x-ray beam and angular dispersive diffraction patterns were collected with a Perkin Elmer detector using an exposure time of 10-20 s. The two dimensional diffraction images were integrated using Fit2D,[28] then analyzed using the JADE software package to extract crystal structure and volume information. To determine the pressure, we used the Vinet equation of state of the copper, with fitting parameters for the bulk modulus $B_0$=133 GPa and its pressure derivative $B_0$'=5.01 from Dewaele et al.[29]

For electrical transport studies under pressure we used an eight-probe designer DAC[30, 31] with 280μm diameter culets, steatite as a pressure-transmitting medium and ruby as the pressure calibrant. [32, 33] A MP35N metal gasket was preindented to an initial thickness of 60 μm, and a 120 μm hole was drilled in the center of the indentation for the sample chamber using an electric discharge machine. A small crystal of $Bi_2Te$ with dimensions of 80-μm diameter and 10-μm thick, was taken out of the larger, micaceous crystallites and placed onto the designer anvil to ensure electrical contact with the tungsten leads exposed on the face of the designer diamond culet. Pressure was measured at room temperature on two separate ruby spheres within the sample chamber in order to estimate pressure distribution across the chamber. Based on previous studies using this type of DAC the error in the pressure at low temperatures was estimated to be 5%[31]. Temperature was measured using a calibrated Cernox thermometer affixed to the outside of the DAC. Electrical transport measurements were made as a function of temperature and magnetic field using the AC Transport option in the Quantum Design PPMS.

## Results

High pressure x ray diffraction studies were performed on $Bi_2Te$ powder in order to explore the universality of the collapse into the high pressure body-centered-cubic (bcc) phase that occurs within the $(Bi_2)_m(Bi_2Te_3)_n$ series. At ambient pressure and temperature the $Bi_2Te$ starts out in the hexagonal phase with the *P-3m* crystal structure, denoted as $Bi_2Te$-I, which persists up to 7 GPa. Upon compression near 5 GPa an intermediate phase begins to appear, denoted as $Bi_2Te$-II. Diffraction peaks belonging to this $Bi_2Te$-II phase are present in the diffraction patterns up to 17 GPa. However, at 9 GPa a second phase transition begins into the $Bi_2Te$-III that completes at 17 GPa. Integrated diffraction patterns for the three phases are shown in figure 1. $Bi_2Te$-II has a very broad pressure range where coexist with either phase I or phase III. We were not able to observe it as a single phase in our room-temperature experiments, and thus we were unable to index the peaks to unambiguously define a crystal structure. The high pressure $Bi_2Te$-III can be indexed to the bcc (*Im-3m*) unit cell with a=3.747 Å at P= 8.8 GPa and it is the stable phase to 47 GPa, the highest pressure attained in this study. Fitting the pressure-volume data points to the Vinet EOS[34, 35] for phase III we obtain $V_o$= 29.88 ($\pm 0.27$)Å, $B_o$=54.2 ($\pm 5$) GPa and $B' = 4.78(\pm 0.32)$. This $Bi_2Te$-III structure is best understood as a disordered substitutional alloy, such that the bismuth and tellurium atoms reside randomly on the (2a) site of the bcc structure with occupancies defined by the 2:1 Bi:Te stoichiometry of the sample. This is believed to be due to the similar atomic radii of the two elements and is further enabled by charge transfer as a result of the application of high pressure. This behavior has also been seen in similar materials such as $(Bi,Sb)_2Te_3$ and $Bi_4Te_3.$ [20, 36-38] Furthermore it seems that the unit cell volume as a function of pressure for the *bcc*-phase in $Bi_2Te$ follows the same P-V curve as do the above mentioned two compounds (fig. 2(a)). We also note that concomitant with the phase transformation from phase I to phase III, there is a volume collapse of approximately 9%. The fact that $Bi_2Te$ also collapses into the bcc phase *strongly* suggests that, despite having disparate ground state phases, the entire $(Bi_2)_m(Bi_2Te_3)_n$ series converges into the bcc phase at high pressures. These structural transformations are ubiquitous across this series, which suggests that there should also be concurrent transformations occurring in the electronic structure as well.

High-pressure magnetotransport measurements were performed to investigate the effects of the high-pressure structural phase transformations on the electronic structure of $Bi_2Te$. Hall resistance at 25K as a function of applied magnetic field is shown in Figure 3 for selected pressures. It can be seen that there is a clear change in the Hall resistance between 5.4 and 7.6 GPa, coincident with the phase transformation from the hexagonal phase ($Bi_2Te$-I) to the intermediate, mixed phase ($Bi_2Te$-II) observed in x-ray diffraction. The linearity of $R_{xy}$ as a function of applied magnetic field suggests a dominant carrier type (fig. 3) for all pressures at T=25K, but for pressures below 9.2 GPa a multiband picture of $Bi_2Te$ emerges in the temperature dependence of the Hall coefficient (fig. 4). At 2.8 GPa there is a crossover from negative to positive Hall coefficient near 70K demonstrating a change in the dominant charge carrier type. From our x-ray diffraction measurements we know that $Bi_2Te$ is in a mixed phase regime above 5 GPa, but as pressure is increased above 2.8 GPa the crossover temperature in the Hall sign decreases, as does the magnitude of the Hall coefficient, suggesting a net increase in the carrier density as a function of pressure. In the mixed phase above 5.4 GPa there is very little change in the carrier concentration, compared to the concentration in the low-pressure, hexagonal phase, $Bi_2Te$-I. This can be clearly seen in figure 2(b), where the slope of the Hall resistance versus applied field, $R_H$, and the longitudinal resistance, $R_{xx}$, are plotted as a function of pressure. The Hall coefficient, $R_H$, is inversely proportional to the carrier density and, if the carrier concentrations are calculated using a single band model, there is an increase in carrier concentration within order of magnitude from $\sim 10^{21}$ cm$^{-3}$ at 2.8 GPa—more typical of a semimetal— to $\sim 10^{23}$ cm$^{-3}$ for pressures of 6.4 GPa and above—more characteristic of a typical metal. Considered within the single band model, this suggests that there is a metallization of $Bi_2Te$ at these higher pressures there should now be more electrons available to form superconducting pairs, as has been seen in many of the other members of the $(Bi_2)_m(Bi_2Te_3)_n$ series[17, 18, 20].

Low-temperature, high-pressure transport data (Fig. 5) shows the onset of superconductivity at 11.5 GPa. Though $Bi_2Te$ did not fully enter the superconducting state by our minimum temperature of 1.8K, we can extrapolate that the midpoint of the transition should have a $T_c \cong 2.1$ K (see fig. 5). At the next pressure, $P = 12.9$ GPa, there are two transitions visible as it enters the superconducting state. This correlates with the fact that the system is in the intermediate $Bi_2Te$-II and $Bi_2Te$-III mixed phase, with a different $T_c$ for the two states (see fig. 2(c)). $T_c$ increases sharply between 11.5-12.9 GPa for both transitions, with $T_{c,II}$= 5.8K and $T_{c,III}$= 7.7K. Once the pressure is increased to P= 14.5 GPa there is a single, sharp superconducting transition, which gives the maximum $T_c$= 8.6 K with a transition width of only 0.57 K. Although from the x-ray diffraction data we know that $Bi_2Te$ is still in the mixed phase at P= 14.5 GPa, the superconducting transition seems dominated by the $Bi_2Te$-III (bcc) phase, and has a $T_c$ similar to the maximum $T_c$ of Bi[17], $Bi_4Te_3$[20] (both 8.4 K), and $Bi_2Te_3$ [18](9.3 K) under pressure, all of which transform into a high-pressure bcc phase. After reaching its maximum, $T_c$ decreases monotonically down to $T_c$ = 4.9 K at 32 GPa.

The universality of the superconducting state of the $(Bi_2)_m(Bi_2Te_3)_n$ series in the high-pressure bcc phase can be seen more clearly by plotting the reduced critical temperature, $T_c/T_{c,max}$ versus $P/P_{c,max}$, where $P_{c,max}$ is the pressure where $T=T_{c,max}$, as is shown in figure 2(c) inset. This shows the linear trend common to the series as $T_c$ is suppressed by increasing pressure after reaching $T_{c, max}$ at a rate $dT_c/dP$= -0.34±0.02 K/GPa. The linear suppression of $T_c$ is consistent with BCS-type phonon-mediated superconductivity in $Bi_2Te$, in which pressure raises the average phonon frequency and reduces the electron-phonon coupling, thus decreasing the pairing mechanism strength. [39, 40] To further investigate the nature of the superconducting state using a treatment consistent with a BCS superconducting state, we can look at the electron-phonon coupling through the McMillan formula to find the volume Grüneisen parameter and compare it with other BCS superconductors.

The McMillan formula is given as $T_c = (\langle\omega\rangle/1.2)\exp([-1.04(1+\lambda)]/[\lambda-\mu^*(1+0.62\lambda)])$, valid in the strong coupling regime ($\lambda \leq 1.5$), and connects $T_c$ with the electron-phonon coupling parameter $\lambda$, an average phonon frequency $\langle\omega\rangle$, and the screened Coulomb repulsion $\mu^*$ (taken here to be 0.1).[41] By taking the logarithmic volume derivative of $T_c$, we get the relation

$$\frac{d \ln T_c}{d \ln V} = -\gamma + \Delta\left\{\frac{d \ln \eta}{d \ln V} + 2\gamma\right\},$$

in which $\gamma = -d \ln\langle\omega\rangle/d \ln V$ is the Grüneisen parameter, $\eta = N(E_f)\langle I^2\rangle$ is the Hopfield parameter, which is given by the product of the electronic density of states and the average-squared electronic matrix element, and $\Delta = 1.04\lambda[1+0.38\mu^*]/[\lambda-\mu^*(1+0.62\lambda)]^2$. [42] Although the average phonon frequency $\langle\omega\rangle$ has not been determined experimentally for $Bi_2Te$ at high pressure, we can test a range of values of 100, 200, and 300K, that cover the range of the typical values. We choose P= 14.5 GPa, where $T_c$ is a maximum at 8.6 K, to look at these values. Substituting $T_c$= 8.6 K, $\langle\omega\rangle$ =100K, and $\mu^*$ =0.1 into the McMillan equation and solving for $\lambda$, we find $\lambda$=1.17. We can then substitute this value into the equation for $\Delta$, which gives a value of $\Delta$=1.28. All of these values can then be substituted into the above equation to find $\gamma$. We can extract the value of $d \ln T_c/d \ln V$ from our experimental data by combining our equation of state data (Fig. 4a) with the superconducting phase diagram (Fig. 4c). If we plot $\ln T_c$ vs $\ln V$ and take the derivative near P= 14.5 GPa, we find a value of $d \ln T_c/d \ln V$ =5.55 K/Å$^3$. Substituting this value into the above equation for $\gamma$, along with our value of $\Delta$, and taking the value of $d \ln \eta/d \ln V$= -1,[43] assuming again that $Bi_2Te$ behaves like a simple p-electron metal, we are able to extract a value of $\gamma$= 4.38. To test the parameters of this calculation we can vary the value of $\langle\omega\rangle$= 200 and 300K, which gives values of $\gamma$= 2.3 and 1.9, respectively, closer to that of tin ($\gamma \approx +2.1$).[44] While yielding a $\gamma$ value closer to that expected for simple metals like Sn, $\langle\omega\rangle$=200 or 300 K are substantially higher values than that of Sn ($\langle\omega\rangle \approx$ 110K).[41] Given the strong compression of the lattice at the pressures where superconductivity is observed, it is likely that there is commensurate phonon stiffening, which would serve to increase <ω>. In light of this potential phonon stiffening, a value of <ω>=200 K does not seem unreasonable, and therefore the associated γ may be a reasonable estimate. Even with uncertainties in <ω>, the γ values calculated for Bi$_2$Te suggest that it behaves like a simple, BCS type phonon-mediated superconductor in the high-pressure, bcc phase. Future experiments studying the high-pressure phonon dispersions in Bi$_2$Te are needed to determine the actual value of <ω> and thereby more accurately determine the nature of the superconducting state in Bi$_2$Te, and possibly the rest of the homologous series.

To investigate the pair-breaking mechanism in greater detail we applied magnetic fields to study how the superconducting state is suppressed as a function of pressure and magnetic field. This is shown for a single pressure of P= 14.5 GPa for applied fields up to 30 kOe (Fig. 6), but was performed at all of the pressures in order to understand the pressure evolution of the superconducting state. Figure 7 shows the rate of suppression of T$_c$ with applied magnetic field at all pressures up to 32.0 GPa. The rate of suppression is quasilinear at all pressures down to the lowest temperatures measured. This quasilinear trend of H$_{c2}$(T) has also been observed in Bi$_4$Te$_3$ [20], Bi$_2$Se$_3$ [26], Cu$_x$Bi$_2$Te$_3$ [45], as well as YPtBi [46, 47] and ErPdBi [48]. Considering that all of these materials contain bismuth, these results suggest that strong spin-orbit scattering is the dominant pair breaking mechanism in these materials.

Within the Werthamer-Helfand-Hohenberg (WHH) model [49] for orbitally limited superconductors we can calculate the upper critical field at P= 14.5 GPa as H$_{c2}$(0)= -0.7 T$_c$ × dH$_{c2}$/dT|$_{T=Tc}$= 1.94 T, where T$_c$(H=0) = 8.66 K taken at the midpoint of the superconducting transition, and dH$_{c2}$/dT =-0.321 T/K . This is lower than the measured H$_{c2}$(2K)= 2.12 T, suggesting that there is another mechanism involved in the superconducting pair breaking. If H$_{c2}$ is Pauli limited, then the superconducting gap energy Δ is equal to the Zeeman energy and therefore the Pauli limiting field H$_P$=1.84$T_c$= 15.93 T at P= 14.5 GPa. Since the measured H$_{c2}$ is higher than the orbital limit but significantly lower than $H_P$, we can assume that both orbital and Pauli limiting effects are in play, with the orbital mechanism being the dominant one. To incorporate both mechanisms into our calculations of $H_{c2}$, we can recalculate $H_{c2}$ using the modified formula, which includes the Maki parameter $\alpha = \sqrt{2} H_{c2}^{orb}/H_p$ [50], so that $H_{c2}^{\alpha} = H_{c2}^{orb}/\sqrt{(1+\alpha^2)} = 1.79$ T at P=14.5 GPa, where $\alpha=0.172$, $H_{c2}^{orb}= 1.94$ T, and $H_P= 15.93$ T. Including the Maki parameter only lowers the calculated $H_{c2}(T)$, providing further evidence that the pair breaking in $Bi_2Te$ is driven by spin-orbit scattering and not Pauli spin-paramagnetism.

To compare our results within the WHH theory for spin-singlet superconductors we plot the normalized critical field $h^*(t)=(H_{c2}/T_c)/|dH_{c2}/dT|_{t=1}$ as a function of the normalized temperature $t=T/T_c$ for selected pressures up to 32 GPa (figure 8). All of the pressures follow the same $h^*(t)$ slope and remain linear down to the lowest temperatures measured of $t\sim0.3$. Based on the strong spin-orbit effects of Bi based materials, we assumed that spin-orbit scattering was the main contributor to the pair-breaking mechanisms in $Bi_2Te$ and therefore we include spin-orbit effects ($\lambda_{so}$) within the WHH formalism, in a similar fashion to reference [51], and compare them to our experimental results (Fig. 8): the blue, dashed line is the WHH in the dirty limit ($\alpha=0$); the red, dashed line includes a value of $\alpha=0.2$ (close to our experimental value of 0.172 calculated for P= 14.5 GPa) and $\lambda_{so}=0$; and the black, dashed line includes $\alpha=0.2$ and an unphysically large $\lambda_{so}=100$. The addition of $\alpha=0.2$ shifts the $h^*(t)$ curve down only slightly, and only for $t\leq0.3$. Because $\alpha$ is so small in $Bi_2Te$, increasing $\lambda_{so}$ barely shifts the $h^*(t)$ curve up even for $\lambda_{so}=100$, a spin-orbit scattering strength, which is well outside the limits of WHH. Considering that our data deviate from all of these models below $t\sim0.5$, it is clear that a theory which includes very strong spin-orbit interactions must be developed to deal with this and similar spin-orbit driven superconductors, as attested to by a growing body of experimental work[26, 45, 47, 52].

In conclusion, we have found that $Bi_2Te$ provides further evidence of a universal behavior of the series to collapse into the bcc structure at high-pressure. Although

Bi$_2$Te enters a high-pressure mixed phase region between roughly 5-17 GPa, we were able to observe a semimetal-metal transition near 5.4 GPa in both linear and Hall resistance measurements. We were able to establish that there is a dominant carrier type below 5.4 GPa via magnetic field and temperature dependence measurements, though it was difficult to unambiguously determine carrier type and concentration in the mixed phase. We also made the first discovery of pressure induced superconductivity in Bi$_2$Te, with a maximum $T_c$=8.66K at P= 14.5 GPa. Bi$_2$Te has a similar maximum $T_c$ (~8K) to other materials in the (Bi$_2$)$_m$(Bi$_2$Te$_3$)$_n$ series, as well as a similar rate of suppression of $T_c$ with pressure. Analysis of the pressure-dependence of the superconducting state suggests that Bi$_2$Te is likely a conventional, BCS superconductor, but the linear suppression of $T_c$ with field implies unconventional and strong spin-orbit scattering pair-breaking effects. This pair-breaking is outside of the scope of the conventional WHH theory and requires a new treatment that is able to incorporate strong spin-orbit scattering. Although establishing whether Bi$_2$Te has topologically nontrivial surface states is outside the scope of this paper, as a member of the homologous series (A$_2$)$_n$(A$_2$B$_3$)$_m$, where A=Bi, Sb, Pb, Ge and B=Te, Se, S, many permutations and dopings of these and similar materials[14, 15, 45, 53-55] have produced TIs, making it likely that Bi$_2$Te is one also. Future work should aim to confirm whether Bi$_2$Te is a TI via theoretical calculations as well as experimental techniques like ARPES, STM or quantum oscillations. This would help to confirm the universality of the infinitely adaptive (Bi$_2$)$_m$(Bi$_2$Te$_3$)$_n$ series, not only with regard to the superconducting state and high-pressure crystal structure, but also in the topological nature of these materials as well.

This work was performed under LDRD (Tracking Code 14-ERD-041) and under the auspices of the US Department of Energy by Lawrence Livermore National Laboratory (LLNL) under Contract No. DE-AC52- 07NA27344. Portions of this work were performed at HPCAT (Sector 16), Advanced Photon Source (APS), Argonne National Laboratory. HPCAT operations are supported by DOE-NNSA under Award No. DE-NA0001974 and DOE-BES under Award No. DE-FG02-99ER45775, with partial instrumentation funding by NSF. The Advanced Photon Source is a U.S.

Department of Energy (DOE) Office of Science User Facility operated for the DOE Office of Science by Argonne National Laboratory under Contract No. DE-AC02-06CH11357. Beamtime was provided by the Carnegie DOE-Alliance Center (CDAC). YKV acknowledges support from DOE-NNSA Grant No. DE-NA0002014.

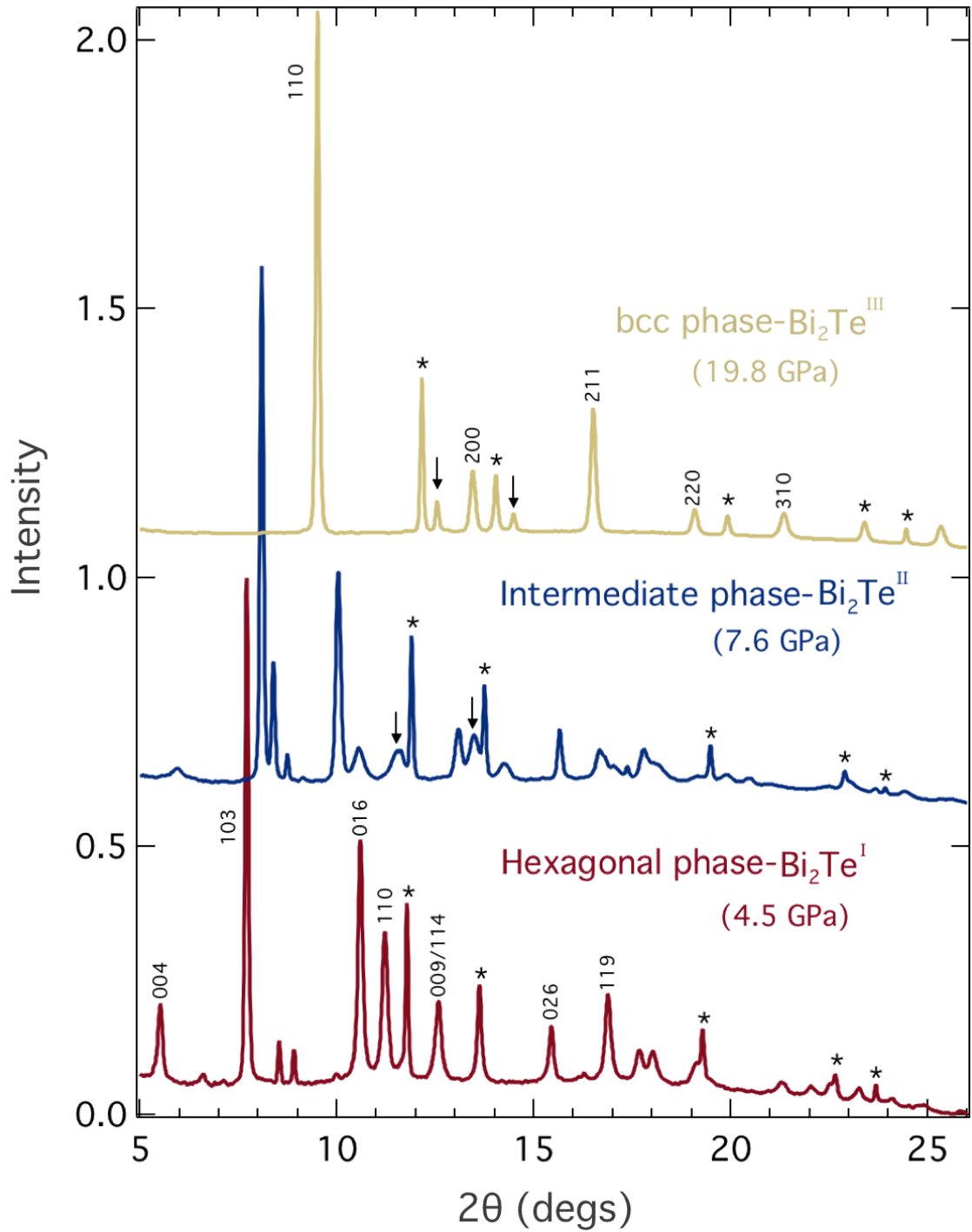

**FIG. 1** (Color online) X-ray diffraction patterns at pressures within the hexagonal (P<5 GPa), intermediate (5 GPa <P<17 GPa) and body-centered-cubic phases (P>17 GPa). The *h k l* indices of some of the diffraction reflections are shown for the hexagonal and bcc phase of $Bi_2Te$. Stars are the peaks from the copper pressure calibrant and the arrows show the peaks due to the neon pressure medium.

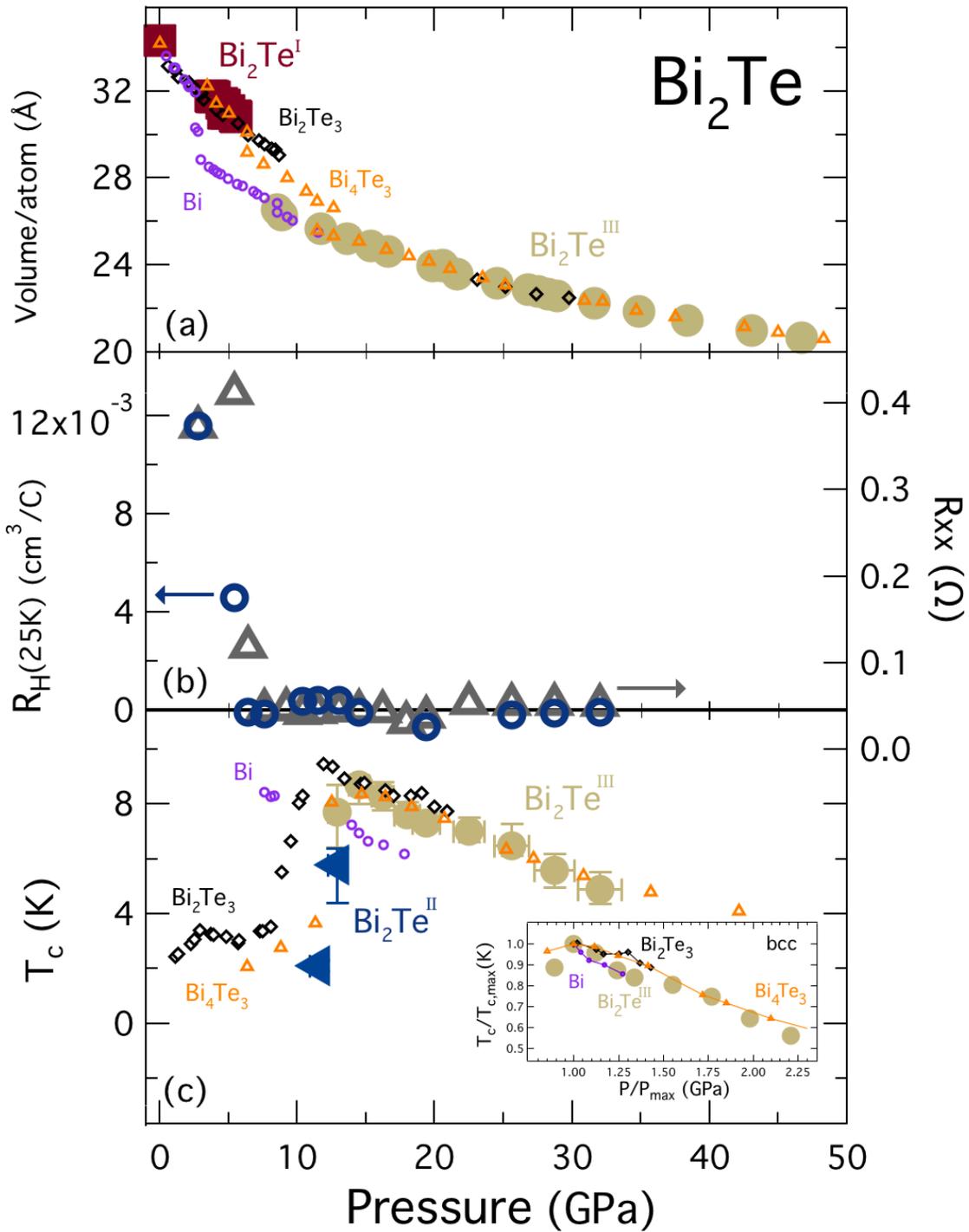

FIG.2 (Color online) (a) Bi$_2$Te shows universal behavior in the high pressure phase as it collapses into the bcc phase, just as the other members of the infinitely adaptive (Bi$_2$)$_m$(Bi$_2$Te$_3$)$_n$ series do. This result confirms this family of compounds, with disparate electronic and crystal structure ground states, can be tuned by the application of pressure to elicit a universal metallic, superconducting phase with the body-centered cubic crystal structure. (b) The electronic phase transitions that accompany the structural transitions are seen in the semimetal to metal crossover near 5 GPa, as well as the change in charge

carrier type between 10-14 GPa. (c) $T_c$ as a function of pressure for $Bi_2Te$ plotted with the other members of the homologous family showing similar trends for peaks of $T_c$ before entering the bcc phase. (c, inset) $T_c$ normalized to maximum $T_c$ in the bcc phase to show the same relation of slope for $T_c$ vs pressure in the bcc phase for the entire $(Bi_2)_m(Bi_2Te_3)_n$ series.

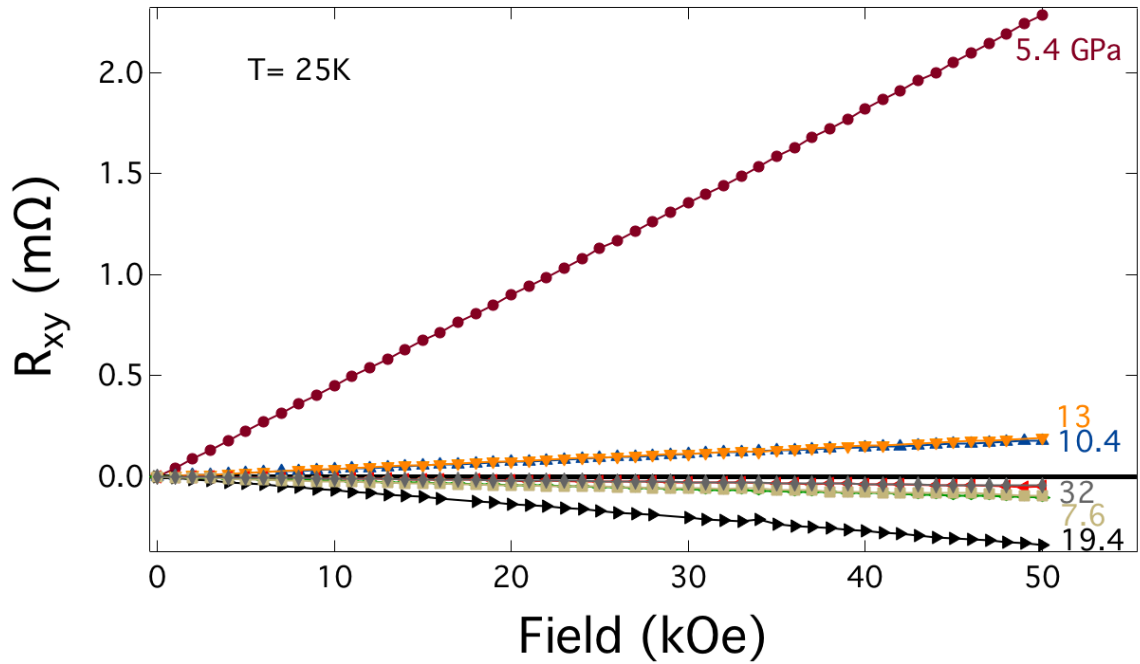

FIG. 3 (Color online) Hall resistance at 25 K as a function of applied magnetic field for pressures between 5.4-32 GPa. There is a significant decrease in the Hall resistance above 5.4 GPa, which indicates a drastic change in the carrier concentration. Also, the change in the sign of the slope indicates that the majority carrier changes as pressure is increased, though the fact $R_{xy}$ vs H remains linear up to 50 kOe suggests that there is a single dominant charge carrier for all of the pressures studied.

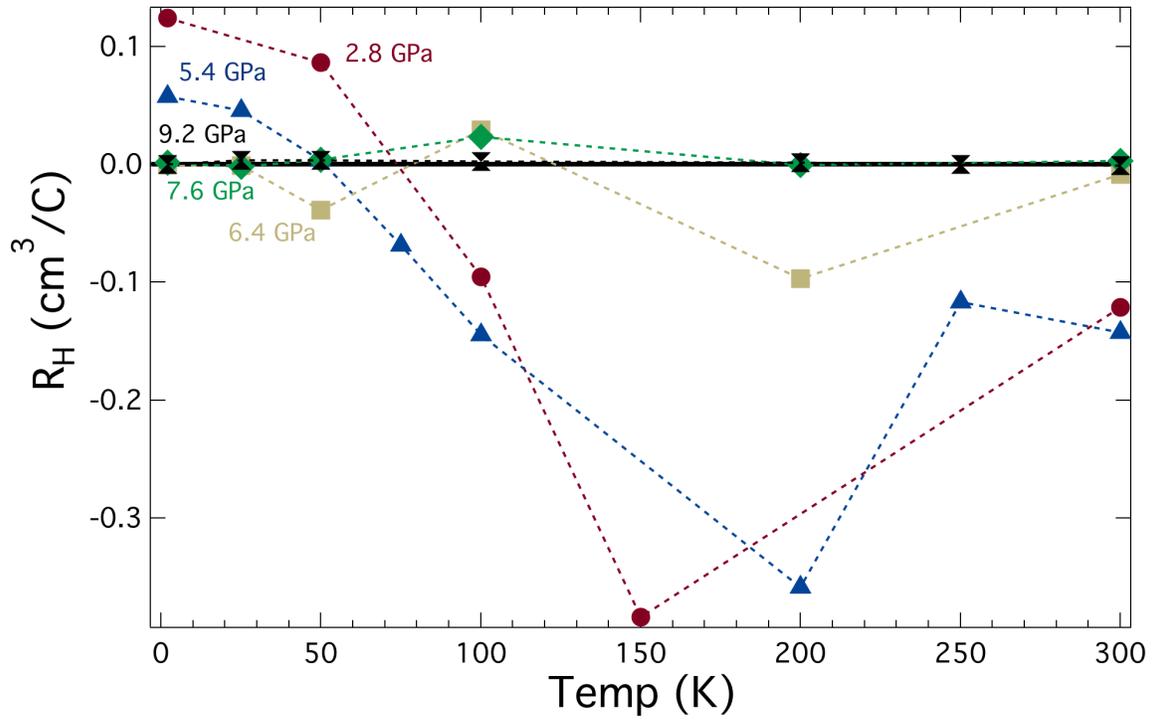

FIG.4 (Color online) Hall coefficient plotted as a function of temperature at increasing pressures shows the suppression of the carrier crossover from near 70K at 2.8 GPa to 50K at 5.4 GPa until it is nearly flat by 7.6 GPa, with no visible change in carrier as a function of temperature. Though it is difficult to interpret these results given that the system is in the intermediate, mixed phase region above 5 GPa, this figure shows that both temperature and pressure (fig. 2b) affect the band structure of $Bi_2Te$ and drive it from a semimetal to a metal via two separate tuning parameters.

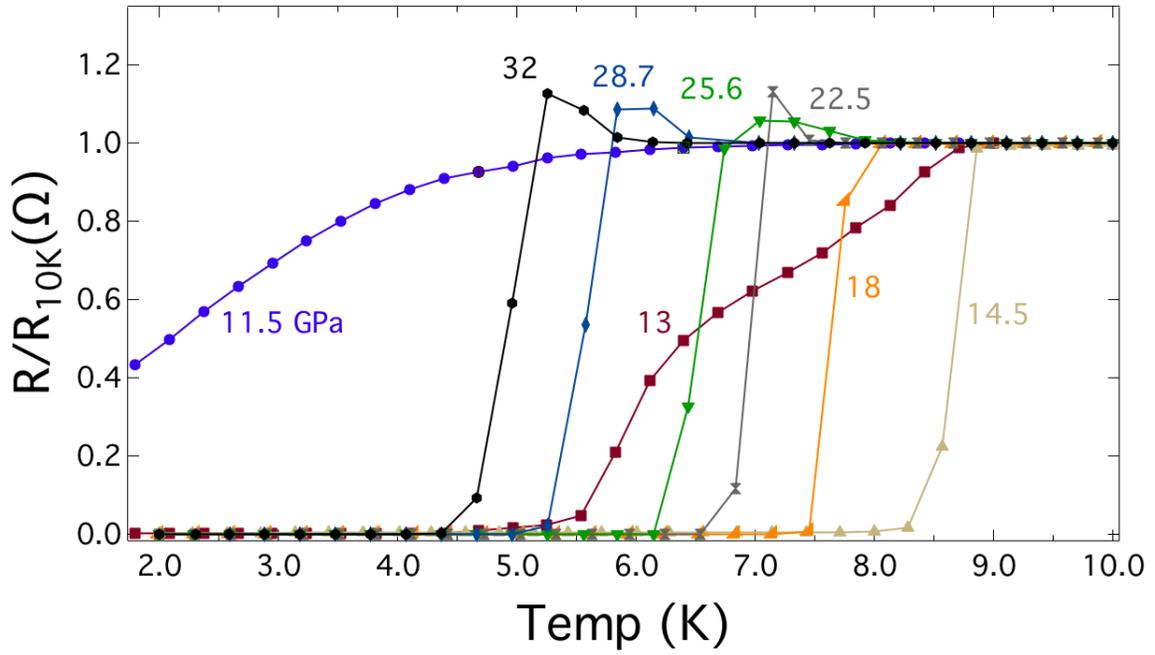

FIG.5 (Color online) Resistance as a function of temperature for pressures from 11.5-32 GPa showing the onset and subsequent suppression of the superconducting state in $Bi_2Te$. The resistance is normalized to the value of the resistance at 10K for each particular pressure just to emphasize the change in $T_c$ as a function of pressure.

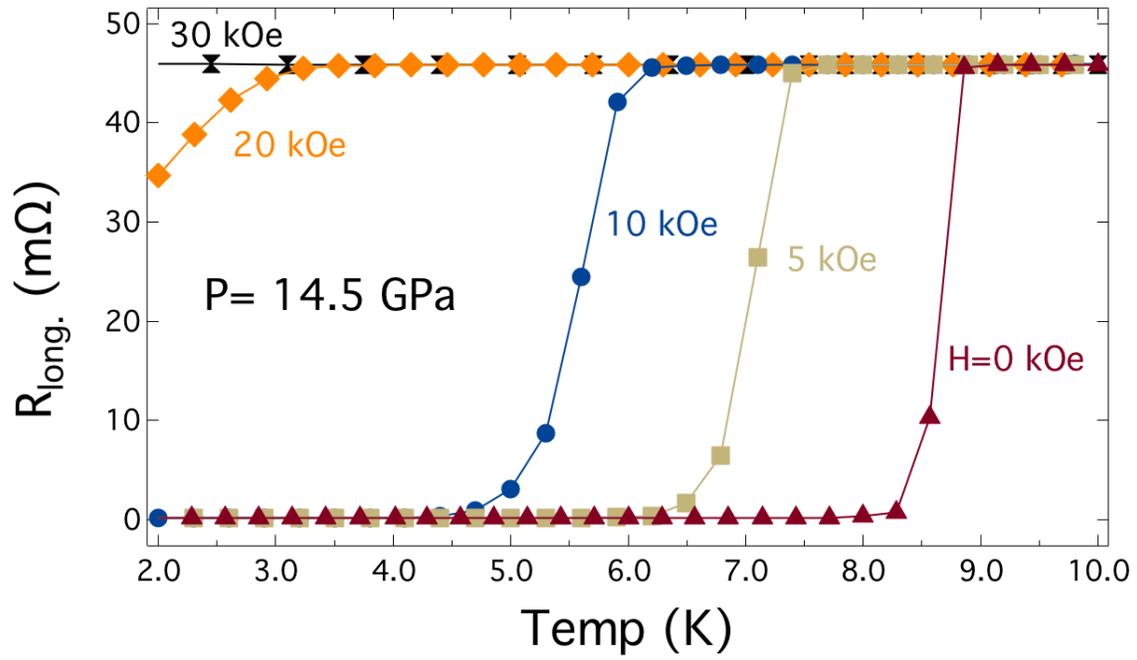

FIG. 6 (Color online) Suppression of the superconducting transition with magnetic field at P=14.5 GPa shown in plots of resistance as a function of temperature. The linear trend of $dT_c/dH$ can be more clearly seen in the phase diagram shown in figure 7.

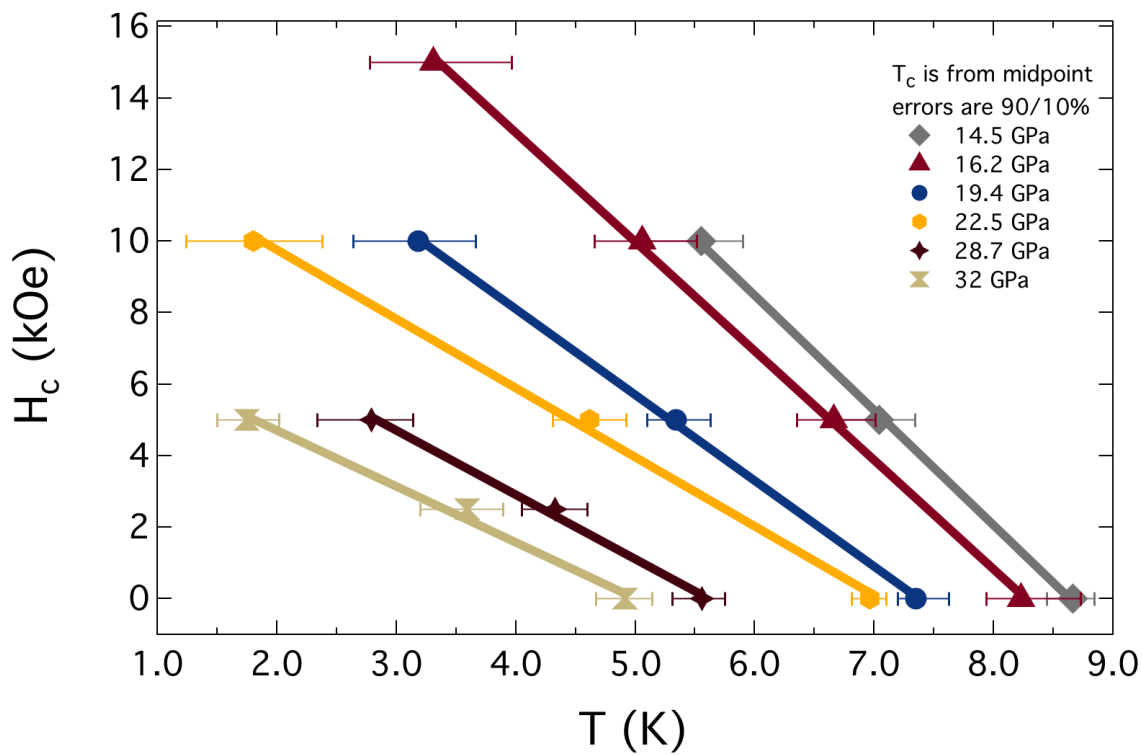

**FIG. 7 (Color online) H$_c$ is suppressed linearly with temperature for all of the pressures studied. Temperature error bars are the values of T$_c$ at the 90% and 10% values of the resistance maximum in the normal state.**

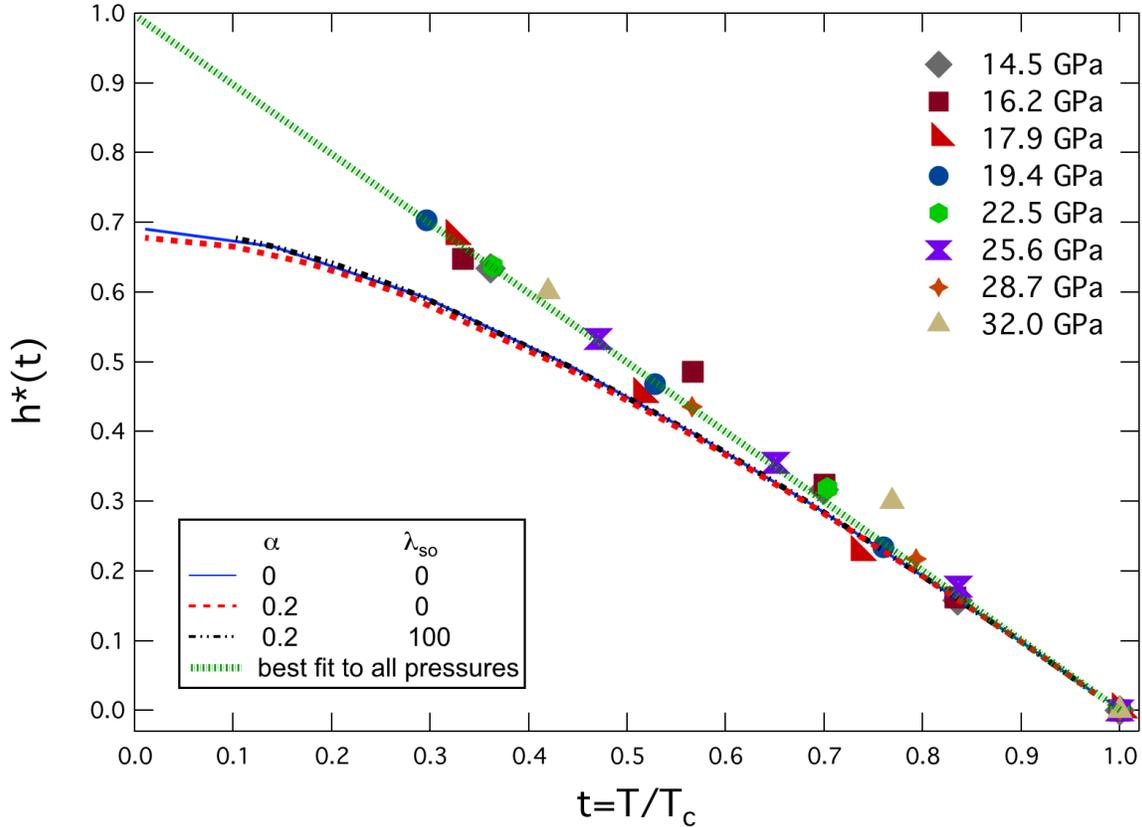

**FIG. 8** (Color online) Graph showing the renormalized $H_{c2}(T)$ for selected pressures using the reduced temperature $T/T_c$, and the reduced critical field (see text) in order to compare with theory. The fitting is done using the classical WHH model in which the blue dashed line is the limit of the Maki parameter, $\alpha=0$ and the spin-orbit scattering strength, $\lambda_{so}=0$. To compare our data with the WHH model we used $\alpha=0.2$ and varied $\lambda_{so}$ to see the effect of spin-orbit scattering. As is shown in the figure, due to the fact that $\alpha$ is so small, even with values of $\lambda_{so}$ up to 100 (well outside of WHH conditions) our data begins to deviate from the models for $t<0.7$.